%% file: main.tex
\documentclass[sigconf]{acmart}
\usepackage{booktabs} 

\setcopyright{rightsretained}
\renewcommand\footnotetextcopyrightpermission[1]{}

\acmPrice{15.00}

\begin{document}
\input{macros}
\title{An Analysis of United States Online Political Advertising Transparency}

\author{Laura Edelson}
\affiliation{
  \institution{New York University}}

\author{Shikhar Sakhuja}
\affiliation{
  \institution{New York University}}

\author{Ratan Dey}
\affiliation{
  \institution{New York University}}

\author{Damon McCoy}
\affiliation{
\institution{New York University}}

\begin{abstract}

During the summer of 2018, Facebook, Google, and Twitter created policies and implemented transparent archives that include U.S. political advertisements which ran on their platforms.
Through our analysis of over \totalAds\ ads with political content, we show how different types of political advertisers are disseminating U.S. political messages using Facebook, Google, and Twitter's advertising platforms. We find that in total, ads with political content included in these archives have generated between \totalMinImpressions\ - \totalMaxImpressions\ impressions and that sponsors have spent over \$300 million USD on advertising with U.S. political content. 

We are able to improve our understanding of political advertisers on these platforms. We have also discovered a significant amount of advertising by quasi for-profit media companies that appeared to exist for the sole purpose of creating deceptive online communities focused on spreading political messaging and not for directly generating profits. Advertising by such groups is a relatively recent phenomenon, and appears to be thriving on online platforms due to the lower regulatory requirements compared to traditional advertising platforms. 


We have found through our attempts to collect and analyze this data that there are many limitations and weaknesses that enable intentional or accidental deception and bypassing of the current implementations of these transparency archives. We provide several suggestions of how these archives could be made more robust and useful. 
Overall, these efforts by Facebook, Google, and Twitter have improved political advertising transparency of honest and, in some cases, possibly dishonest advertisers on their platforms. We thank the people at these companies who have built these archives and continue to improve them. 

\end{abstract}

%
%



\maketitle

\input{body.tex}

\bibliographystyle{ACM-Reference-Format}
\bibliography{bibliography}

\end{document}

%% file: macros.tex
\newcommand{\totalAds}{1.3 million}
\newcommand{\totalMinImpressions}{8.67 billion}
\newcommand{\totalMinSpend}{181 million}
\newcommand{\totalMaxImpressions}{33.8 billion}
\newcommand{\totalMaxSpend}{613 million}
\newcommand{\totalTopSpendOne}{1}
\newcommand{\totalTopSpendOneNumber}{1}
\newcommand{\totalTopSpendTwo}{1}
\newcommand{\totalTopSpendTwoNumber}{1}
\newcommand{\totalTopSpendThree}{1}
\newcommand{\totalTopSpendThreeNumber}{1}

\newcommand{\overallFBAds}{1.26 M}
\newcommand{\overallFBAdSponsors}{24 K}
\newcommand{\overallFBPages}{38 k}
\newcommand{\overallFBMinImpressions}{7.35 B}
\newcommand{\overallFBMaxImpressions}{21.12 B}
\newcommand{\overallFBMinSpend}{135 M}
\newcommand{\overallFBMaxSpend}{567 M}
\newcommand{\overallFBFirstDate}{July 14$^{th}$, 2014}
\newcommand{\overallFBLastDate}{October 21$^{st}$, 2018}
\newcommand{\preArchiveFBAds}{13.5 K}
\newcommand{\averageFBImpressions}{6 K}
\newcommand{\averageFBMaxImpressions}{17 K}
\newcommand{\averageFBSpend}{107}
\newcommand{\averageFBMaxSpend}{450}

\newcommand{\overallGoogleAds}{41 K}
\newcommand{\overallGoogleAdSponsors}{616}
\newcommand{\overallGoogleMinImpressions}{1.3 B}
\newcommand{\overallGoogleMaxImpressions}{11.6 B}
\newcommand{\overallGoogleMinSpend}{11 M}
\newcommand{\overallGoogleMaxSpend}{368 M}
\newcommand{\overallGoogleSpend}{45 M}
\newcommand{\overallGoogleFirstDate}{May 31st, 2018}
\newcommand{\overallGoogleLastDate}{October 21$^{st}$, 2018}
\newcommand{\averageGoogleImpressions}{32 K}
\newcommand{\averageGoogleMaxImpressions}{283 K}
\newcommand{\averageGoogleSpend}{1 K}

\newcommand{\overallTwitterAds}{1808}
\newcommand{\overallTwitterAdSponsors}{88}
\newcommand{\overallTwitterImpressions}{118 M}
\newcommand{\overallTwitterSpend}{1.6 M}
\newcommand{\overallTwitterFirstDate}{December 21$^{st}$, 2016}
\newcommand{\overallTwitterLastDate}{October 21$^{st}$, 2018}
\newcommand{\preArchiveTwitterAds}{25}
\newcommand{\averageTwitterImpressions}{65 K}
\newcommand{\averageTwitterSpend}{885}

\newcommand{\overallProPublicaAds}{81,052}
\newcommand{\postArchiveProPublicaAds}{33,308}
\newcommand{\overallProPublicaPages}{2,363}
\newcommand{\overallProPublicaAdSponsors}{2395}
\newcommand{\overallProPublicaFirstDate}{July 31st, 2017}
\newcommand{\overallProPublicaLastDate}{October 18th, 2018}
\newcommand{\propublicaAdsInArchive}{18,010}
\newcommand{\propublicaAverageSpend}{644}

\newcommand{\totalFBAdsFC}{161 K}
\newcommand{\totalFBAdvertisersFC}{1 K}
\newcommand{\totalFBImpressionsFC}{800 M}
\newcommand{\totalFBMaxImpressionsFC}{2.4 B}
\newcommand{\totalFBSpendFC}{12 M}
\newcommand{\totalFBMaxSpendFC}{60 M}
\newcommand{\averageFBImpressionsFC}{5 K}
\newcommand{\averageFBMaxImpressionsFC}{15 K}
\newcommand{\averageFBSpendFC}{74}
\newcommand{\averageFBMaxSpendFC}{373}

\newcommand{\totalGoogleAdsFC}{15 K}
\newcommand{\totalGoogleAdvertisersFC}{534}
\newcommand{\totalGoogleImpressionsFC}{280 M}
\newcommand{\totalGoogleMaxImpressionsFC}{3 B}
\newcommand{\totalGoogleSpendFC}{13.5 M}
\newcommand{\averageGoogleImpressionsFC}{18 K}
\newcommand{\averageGoogleMaxImpressionsFC}{183 K}
\newcommand{\averageGoogleSpendFC}{850}

\newcommand{\totalTwitterAdsFC}{1 K}
\newcommand{\totalTwitterAdvertisersFC}{54}
\newcommand{\totalTwitterImpressionsFC}{100 M}
\newcommand{\totalTwitterSpendFC}{1.4 M}
\newcommand{\averageTwitterImpressionsFC}{95 K}
\newcommand{\averageTwitterSpendFC}{1329}

\newcommand{\fbUnvettedSponsorAds}{96,106}
\newcommand{\fbUnvettedSponsorSpend}{42.8 million}
\newcommand{\fbUnvettedSponsorImpressions}{670 million}

\newcommand{\overallCategorizedAdSponsorCount}{12,833}
\newcommand{\overallCategorizedAdCount}{907,840}

\newcommand{\categorizedFBAdSponsorCount}{12,192}
\newcommand{\categorizedFBAdSponsorAdCount}{}
\newcommand{\governmentCountFB}{14}
\newcommand{\governmentAdCountFB}{1,656}
\newcommand{\individualCountFB}{7}
\newcommand{\individualAdCountFB}{8,216}
\newcommand{\politicalPartyCountFB}{724}
\newcommand{\politicalPartyAdCountFB}{57,177}
\newcommand{\unionCountFB}{27}
\newcommand{\unionAdCountFB}{8,892}
\newcommand{\forProfitCountFB}{399}
\newcommand{\forProfitAdCountFB}{38,533}
\newcommand{\PACCountFB}{849}
\newcommand{\PACAdCountFB}{200,694}
\newcommand{\nonProfitCountFB}{807}
\newcommand{\nonProfitAdCountFB}{180,102}
\newcommand{\candidateCountFB}{9244}
\newcommand{\candidateAdCountFB}{331,964}

\newcommand{\categorizedFBPartisanLeanAdSponsorCount}{849}
\newcommand{\categorizedFBPartisanLeanAdSponsorAdCount}{764,006}
\newcommand{\categorizedFBPartisanLeanLeftAdCount}{367,658}
\newcommand{\categorizedFBPartisanLeanRightAdCount}{271,832}
\newcommand{\categorizedFBPartisanLeanNeutralAdCount}{124,516}

\newcommand{\categorizedFBAdTypeCount}{884,213}
\newcommand{\categorizedFBAdTypeDonateCount}{141310}
\newcommand{\categorizedFBAdTypeCommercialCount}{144169}
\newcommand{\categorizedFBAdTypeConnectCount}{258238}
\newcommand{\categorizedFBAdTypeMoveCount}{33234}
\newcommand{\categorizedFBAdTypeInformCount}{400166}

\newcommand{\categorizedGoogleAdSponsorCount}{587}
\newcommand{\categorizedGoogleAdSponsorAdCount}{33,571}
\newcommand{\governmentCountGoogle}{0}
\newcommand{\governmentAdCountGoogle}{0}
\newcommand{\individualCountGoogle}{0}
\newcommand{\individualAdCountGoogle}{0}
\newcommand{\politicalPartyCountGoogle}{3}
\newcommand{\politicalPartyAdCountGoogle}{49}
\newcommand{\unionCountGoogle}{0}
\newcommand{\unionAdCountGoogle}{0}
\newcommand{\forProfitCountGoogle}{5}
\newcommand{\forProfitAdCountGoogle}{1823}
\newcommand{\PACCountGoogle}{35}
\newcommand{\PACAdCountGoogle}{17,310}
\newcommand{\nonProfitCountGoogle}{7}
\newcommand{\nonProfitAdCountGoogle}{723}
\newcommand{\candidateCountGoogle}{537}
\newcommand{\candidateAdCountGoogle}{15,307}

\newcommand{\categorizedGooglePartisanLeanAdSponsorCount}{108}
\newcommand{\categorizedGooglePartisanLeanAdSponsorAdCount}{26,667}
\newcommand{\categorizedGooglePartisanLeanLeftAdCount}{13,133}
\newcommand{\categorizedGooglePartisanLeanRightAdCount}{12,791}
\newcommand{\categorizedGooglePartisanLeanNeutralAdCount}{743}

\newcommand{\categorizedGoogleAdTypeCount}{22,612}
\newcommand{\categorizedGoogleAdTypeDonateCount}{6,914}
\newcommand{\categorizedGoogleAdTypeCommercialCount}{24}
\newcommand{\categorizedGoogleAdTypeConnectCount}{5,708}
\newcommand{\categorizedGoogleAdTypeMoveCount}{308}
\newcommand{\categorizedGoogleAdTypeInformCount}{14,302}

\newcommand{\categorizedTwitterAdSponsorCount}{54}
\newcommand{\categorizedTwitterAdSponsorAdCount}{1808}
\newcommand{\governmentCountTwitter}{0}
\newcommand{\governmentAdCountTwitter}{0}
\newcommand{\individualCountTwitter}{3}
\newcommand{\individualAdCountTwitter}{9}
\newcommand{\politicalPartyCountTwitter}{1}
\newcommand{\politicalPartyAdCountTwitter}{26}
\newcommand{\unionCountTwitter}{0}
\newcommand{\unionAdCountTwitter}{0}
\newcommand{\forProfitCountTwitter}{5}
\newcommand{\forProfitAdCountTwitter}{23}
\newcommand{\PACCountTwitter}{20}
\newcommand{\PACAdCountTwitter}{725}
\newcommand{\nonProfitCountTwitter}{4}
\newcommand{\nonProfitAdCountTwitter}{25}
\newcommand{\candidateCountTwitter}{54}
\newcommand{\candidateAdCountTwitter}{1056}

\newcommand{\categorizedTwitterPartisanLeanAdSponsorCount}{54}
\newcommand{\categorizedTwitterPartisanLeanAdSponsorAdCount}{1808}
\newcommand{\categorizedTwitterPartisanLeanLeftAdCount}{1541}
\newcommand{\categorizedTwitterPartisanLeanRightAdCount}{250}
\newcommand{\categorizedTwitterPartisanLeanNeutralAdCount}{76}

\newcommand{\categorizedTwitterAdTypeCount}{1015}
\newcommand{\categorizedTwitterAdTypeDonateCount}{242}
\newcommand{\categorizedTwitterAdTypeCommercialCount}{0}
\newcommand{\categorizedTwitterAdTypeConnectCount}{223}
\newcommand{\categorizedTwitterAdTypeMoveCount}{200}
\newcommand{\categorizedTwitterAdTypeInformCount}{350}

\newcommand{\studyEndDate}{October 21$^{st}$, 2018}

\newcommand{\averagePagesPerAdvertiserFB}{1.6}
\newcommand{\averagePagesForProfitMedia}{3.2}

\newcommand{\FacebookReportAds}{1.67M}
\newcommand{\FacebookReportSpend}{256M}
\newcommand{\FacebookReportPages}{78K}

%% file: body.tex
\section{Introduction}
Online advertising plays an increasingly important role in political elections. As part of the 2016 U.S. national elections there were a number of controversies regarding an ad-driven propaganda campaign to influence elections~\cite{IRA} and privacy violations~\cite{CambridgeAnal}. In response to these controversies, Facebook, Google, and Twitter have all created policies and implemented products to make transparent and archive U.S. political advertisements that have run on their platforms. A report by Upturn points out in their analysis of Facebook's then-proposed political transparency archive plans that providing effective transparency of only political ads can be tricky in the face of a complex online ad network~\cite{upturn}.

In this paper, we analyze Facebook~\cite{FBarchive}, Google~\cite{Google_Political}, and Twitter's~\cite{Twitter_Data} political ad transparency implementations along with the political ad data included in each archive. Through the lens of these transparency efforts, we perform what is to our knowledge the first large-scale analysis of U.S. online political advertising. Our analysis enables us to begin to understand and describe the parts of online political advertising in the U.S. that have been made accessible and transparent. We collected as much data as was possible from these archives, collecting 75\% of ads archived by Facebook and 100\% of ads archived by Twitter and Google from May, 2018 -- October 21$^{st}$, 2018. In total, we collected and analyzed over \totalAds\ political ads from over 24 thousand sponsors. We additionally connect this data to a public dataset published by ProPublica which was gathered by Facebook users via a browser plugin. The ProPublica dataset provides partial information on how Facebook ads have been targeted to the user seeing them. 

Our analysis was hampered by our inability to collect all of the ads in Facebook's transparency archive due to limitations of their current beta API. It was also hampered by Facebook and Google releasing ranges instead of exact impression data. It is also unclear if spend is the best metric for measuring the impact of an online political advertisement. Thus, there is some level of uncertainty in much of our analysis especially that related to Facebook's platform. We acknowledge that all three of these political advertising transparency archives were rapidly deployed and this has caused some of the issues with what and how it was released. We have worked with Facebook to improve access to the ads in their transparency archive and we hope to work with Google and Twitter so that they can include more political advertisers in their transparency archives.

Given these limitations and biases of our data, we perform an initial large-scale analysis of U.S. online political advertising. As part of this study, we provide information about the audience size of individual political ads based on impression data for each of the platforms. We find that across all three platforms the majority of political ads are small, costing their sponsors less than \$100 USD with 82\% of all Facebook political ads costing between \$0 - \$99. This confirms and quantifies the prevalence of small likely highly targeted ads that can contain custom political messaging. We also create a taxonomy of political ad types based on their intent (i.e., connect, donation, inform, move) and an effective ad type classification methodology based on labeling URLs included in political ads. Using our methodology, we are able to provide a longitudinal analysis of political ads based on the type of ad. Finally, we are able to identify many likely dishonest advertisers that are not correctly disclosing or are obfuscating the real ad sponsor. We categorize these types of advertisers into quasi for-profit media companies and corporate astroturfers. We discuss the limitations and weaknesses that enable intentional or accidental deception and bypassing of the current implementations of these transparency archives. Based on our experiences analyzing these transparency archives, we provide several suggestions of how these archives could be made more robust and useful. 

We have made as much of the data that we have collected as possible public as well as all of our labeling, data collection, and analysis scripts public. Our initial reports have been the basis for many journalists' stories which have improved online political advertising transparency.

\section{Background}

Facebook, Google, and Twitter have all deployed political advertising transparency archives. However, each of these transparency archives have different criteria for the inclusion of ads and different modes of access which we create a taxonomy of in Table~\ref{tab:transparency_implementations}. Each of these transparency archive implementations has strengths and weaknesses and there is currently no ``best'' archive. 

\begin{table}[ht]
    \centering
    \begin{tabular}{l|l|l|l}
          & Facebook & Google & Twitter \\ \hline
        Ads & All Candidates, & Federal  & Federal \\ 
		Included & Issue ads & candidate  &    candidate,    \\
		& & related& Issue ads\\ \hline
        Sponsor & Name & Name, & Name, \\ 
		Info     &      & FEC/EIN & billing info \\ \hline
        Ad Contents & Ranges & Ranges & Exact \\ \hline
        Viewed & Gender, age, & N/A & Gender, age,\\ 
		Audience & geolocation  &      & geolocation \\ \hline
        Targeting & N/A & Age, gender, & Age, gender,\\
        Info      &     & geolocation & geolocation \\ 
                  &     & keywords    &     \\ \hline
        Data & Portal, & Portal, & Portal  \\ 
		Availability  & API (w/NDA) & Database & API \\
    \end{tabular}
    \caption{Transparency Implementations}
    \label{tab:transparency_implementations}
\end{table}

\noindent\textbf{What is an ad?} For purposes of this study, an ad is a record in an archive with a unique id assigned by the platform. While each platform has slightly different information associated with each record, each has 3 categories of information associated with it: content, context, and results. The contents of an ad consist of any text, images, and/or videos seen by an ad viewer. The context of the ad is the information specified by the advertiser about how, when, where, and by whom they want the ad to be seen and how much they are willing to spend for the ad to be seen. Both the content and context of the ad are specified by the advertiser at the time of ad creation. The results of the ad consist of information about who ultimately interacted with the ad and when those interactions happened, and how much was ultimately spent on the ad. Not all of this information is made available about each record in any of the archives, but all of this information is made available by at least one archive. 

 \noindent\textbf{Archive Completeness.} All archives contain a unique id for each ad, impression counts and amount spent for each advertisement, as well as the dates the ad was active. Facebook and Google release ad impression and spend in ranges and Twitter releasing these as exact numbers. Twitter's implementation of releasing exact impression and spend information from every ad enables us to more precisely measure political advertising on their platform. It would help remove uncertainty if Facebook and Google would also release exact ad impressions and spend amounts instead of ranges. We also note that impressions is an imperfect metric to measure the ``reach'' of an ad and it would be useful to also include click and other interaction metrics recorded by the platforms. Facebook and Twitter includes the ad text, image, and video content; Google's archive contains a link to a webpage where this content is viewable, but does not contain the content itself. Twitter is the only platform to release targeting information, including whether users were targeted by geography, age, or gender. Facebook and Google have detailed information about how users are targeting for each ad, based on advertiser produced lists of personally identifiable information, groups they belong to, demographic or income information that the platform has about the user, geography, or keyword search information, however they do not currently make any of this information available in their archives.

\noindent\textbf{Ads Included.} Facebook has the most inclusive policy and includes in their archive ads that meet any of the following criteria:
``(1) Is made by, on behalf of, or about a current or former candidate for public office, a political party, a political action committee, or advocates for the outcome of an election to public office; (2) Relates to any election, referendum, or ballot initiative, including "get out the vote" or election information campaigns; (3) Relates to any national legislative issue of public importance in any place where the ad is being run;~\cite{FB_National} (4) Is regulated as political advertising.''~\cite{FB_political} Facebook allows advertisers to opt into including their ads in the archive. In order to enforce their policy, Facebook uses a combination of user reports and machine learning algorithms to ``catch'' political ads where the sponsor did not opt into making them transparent. 

Google is only including "ads related to elections or issues that feature a federal candidate or officeholder"~\cite{Google_Political}. Google has stated that they plan on expanding the set of ads included in their archive. It is unclear how Google enforces their policy.

Twitter's original policy was limited to only including ads sponsored directly by federal candidates. However, Twitter has since expanded their policy to include: ``(1) Ads that refer to an election or a clearly identified candidate.  (2) Ads that advocate for legislative issues of national importance. A clearly identified candidate refers to any candidate running for federal, state, or local election~\cite{Twitter_Political}.'' Based on our analysis, it appears that Twitter is currently not enforcing their policy well.

\noindent\textbf{Sponsors' Info.}
Facebook only displays a text string which the sponsor provides and is intended to identify who is paying for the ad. We went through the vetting process of becoming a political advertiser on Facebook's platform. This entailed uploading a U.S. identification card which was approved approximately five minutes later, at which time we could start posting political ads. Facebook also validates the address of the advertiser by sending them a post card which must be replied to within 30 days or else the advertiser will be suspended. However, during this 30 day grace period political advertisers can post ads without validating their address.  We disclosed to Facebook and a subsequent independent experiment by Turton, a reporter from Vice news, showed that Facebook does not currently vet this text string which allowed the reporter to post ads appearing to be from U.S. Senators~\cite{vice}. This is a security issue that Facebook has acknowledged but claims there is no effective and scalable vetting techniques~\cite{Merrill}. We will describe the issues that we found with this self-reported text string later in this paper.

Google provides both a text string and a Federal Election Candidate (FEC) ID or EIN (U.S. Tax ID) which is vetted for every political advertiser in their archive. Twitter provides a text string and, when available, the billing info of the political sponsors. Twitter initially did not vet this information but started vetting sponsor's EIN similar to Google on September 30$^{th}$, 2018. Providing a consistent and easy to reference identifier such as FEC or EIN for each sponsor enables us to better study sponsors in Google and Twitter's archive.

\noindent\textbf{Viewed Audience.} Facebook and Twitter both include break downs on the impression viewing audience by gender, age range, and state level geolocation. Google provides this information as a heat map image that we cannot currently extract this information from for our analysis thus we mark Google as ``N/A'' for this category. Google has replied that they will work on releasing this information in a format that we can analyze.

\noindent\textbf{Targeting Info.} Facebook does not include any explicit targeting information in their archive. Google and Twitter makes transparent Age, gender, and geolocation based targeting but do not appear to release other types of targeting criteria, such as audience and content, which are allowed by their advertising platforms. Google also makes transparent some aggregated keyword targeting data. All of the platforms only release partial targeting information, at best, which obscures a key facet of online political advertising.

\noindent\textbf{Data Availability.}
Facebook initially only provided a keyword based portal that was designed for small-scale interactive user exploration of the ads in the archive. Facebook enabled anti-scraping functionality in July 2018 that makes it difficult to collect large-scale data by scraping this portal. In September 2018 Facebook released an API that is currently in beta testing which we have access to after signing an NDA stipulating that we will not publicly release raw data collected from the API. This effectively means that only a small set of U.S. organizations participating in the API beta test have large-scale data collection access to ads in the Facebook archive.

Twitter has provided an open API and list of all accounts included in their transparency archive which allows us to effectively collect all ads included in their archive. Google implemented a portal similar to Facebook's and also releases a BigQuery (SQL-like) database of all the ads included in their archive which is updated weekly. For our use case of large-scale data analysis this database format is ideal.

\section{Data Collection Methodology}

\subsection{Facebook}
Initially we scraped Facebook's archive using a list of keywords that included elected positions (i.e., governor, judge, senator), U.S. state names, and key political issues (i.e., health care, immigration, taxes). Around the end of July 2018, Facebook implemented anti-scraping measures which blocked our scraper. Thus, we had no viable means of collecting large-scale ad data until Facebook implemented their API in September 2018. We have publicly released a report and all of the data that we collected by scraping Facebook's archive user portal before our scraper was blocked~\cite{prior_facebook_ad_study}.

We are part of Facebook's Political Ad Archive API beta testing program~\cite{FB_Api} which allows us to query Facebook's Political Ad Archive for specific keyword terms which is matched against the page name, the disclaimer, or ad text.  Ads returned by Facebook's API are ordered using a proprietary ranking algorithm that was not described to us how it functions. However, most advertisements appear to be returned in chronological order. A single query to Facebook's API returns at most 1,000 ads and we can page through to collect additional ads using pagination functionality as part of the API. Currently there is a limitation in Facebook's Political Ad Archive API beta that prevents us from paging past 8,000 ads. This is problematic because many searches will return far more than 8,000 results.

Information on spend and impressions per ad is only available in broad ranges. For impressions, the ranges presented are: 0 - 999, 1,000 - 4,999, 5,000 - 9,999, 10,000 - 49,999, 50,000 - 99,999, 100,000 - 199,999, 200,000 - 499,999, 500,000 - 999,999. For spend, the ranges presented are: 0 - 99, 100 - 499, 500 - 999, 1,000 - 4,999, 5,000 - 9,999, 10,000 - 49,999, 50,000 - 99,999, 100,000 - 199,999, 200,000 - 499,999, 500,000 - 999,999.

Additionally, the API has very low rate limits. We have found that functionally, we could make at most 3 requests per minute on average before hitting these rate limits. Our goal was to create as comprehensive and representative a dataset as possible. Given the very low rate limits and limits on the number of responses for a given search, our approach was to search by advertising page as much as possible in order to reduce the bias in our data. We are currently able to keep up with the rate of new advertisements appearing in Facebook's political ad transparency archive. We cannot publicly release the raw data that we have collected from Facebook's API due to the agreement that we have signed with Facebook as a requirement for access to their API.

We created a separate approach for discovering pages that are linked to sponsored political ads. Our approach to discovering pages involved scraping Facebook's Political Ad Archive user portal interface. We chose a scraping method for page discovery since our access to Facebook's API is highly rate limited and it would be logistically infeasible to perform the queries required for both page discovery and to collect ads using our API access. Our list was not a complete list of advertisers using Facebook's platform since it depends on good coverage based on our keyword searches. 

Facebook started publishing a comprehensive list after the cutoff for our data analyzed in this study. Our data collection from Facebook's archive is our best effort and was incomplete based on analysis of what is contained in Facebook's transparency report. We have changed our data collection methodology moving forward to discover Facebook pages running political ads using Facebook's weekly transparency reports. This combined with improvements we requested and Facebook implemented to their API after the data collection period for this study will improve our coverage to a mostly complete set of U.S. political ads Facebook has included in their transparency archive.

\subsection{Google}

Google published their archive as a public dataset in a BigQuery (SQL-like) format and committed to keeping it public. However, we observed that ad spend and impression values for ads, and occasionally advertiser information were being changed after their entry into the archive. For this reason, we created separate archives of the dataset on a weekly basis. Additionally, ad text information was not available in the dataset itself, but was viewable at a summary page for each ad. We scraped all of these associated pages to collect ad text to be associated in our dataset with the underlying ad data. Unfortunately, many of these summary pages did not render correctly, so we were only able to collect ad text for approximately 66\% of pages which contained it. Two separate issues prevented collection. First, some ad pages display the message:

\noindent\textit{``Advertisers are able to use approved third party vendors to serve ads on Google. While we are able to review these ads for compliance with advertising policies, due to technical limitations, we are currently unable to display the content of the ad in the Transparency Report.''}

Second, some ad pages displayed the message:

\noindent\textit{``Policy violation
This ad violated Google's Advertising Policy.''}
We recommend that Google change their implementation so that ads served by third party vendors or which were deleted for compliance reasons are still accessible through their transparency archive. Google can place a click-through disclaimer to avoid accidental exposure to policy violated content similar to what Facebook has implemented for deleted advertisements.

Information on spend and impressions per ad are only available in broad ranges. For impressions, the ranges presented are: "<= 10k", "10k-100k", "100k-1M", "1M-10M", "> 10M". For spend, the ranges presented are: "< 100", "100-1k", "1k-50k", "50k-100k", "> 100k".
 The ads in this dataset are a combination of text-only ads that are displayed alongside Google search results and image or video-only ads that are displayed as banner or sidebar ads on Google's AdSense network.
     
In addition to per-ad data, Google also published some aggregate data on a per-advertiser and geographic basis. One of these aggregations was exact weekly spend per advertiser. Throughout this paper, we present minimum numbers for impressions and spend because both Google and Facebook publish ranges for impressions and spend for each ad, instead of exact numbers. These total aggregations, give us a sense of how much error there is when we use these minimum estimates for Google. According to Google, advertisers spent \$\overallGoogleSpend\  on political ads but our minimum estimate of spending was only \$\overallGoogleMinSpend.

\subsection{Twitter}

Twitter publishes a list of all political campaigning advertisers~\cite{twitterAds} which we scrape daily to discover new political campaign advertisers' Twitter accounts. In addition to this list provided by Twitter, we have also manually attempted to identify every federal election candidates' personal or campaign Twitter account. We then query each account daily using Twitter's API perform to collect updated information on all promoted tweets and detect federal election candidates which are not listed on Twitter's political campaigning advertisers page but are sponsoring tweets. During our scraping, we have noticed that some promoted tweets were deleted and are replaced with the text:

\noindent\textit{``This Tweet is not available because it includes content that violated Twitter Ads Policies.''}

The information for these deleted promoted tweets is no longer accessible through Twitter's political transparency archive. However, if we have scraped them before they were deleted we have retained the content and information about these promoted tweets. We recommend that Twitter change their implementation so that promoted tweets which were deleted are still accessible through their transparency archive. Twitter can place a click-through disclaimer to avoid accidental exposure to policy violated content similar to what Facebook has implemented for deleted advertisements. We have made public all of the data that we have collected from Twitter's transparency archive.

Additionally, we noted that there were several accounts of federal candidates that were not being archived according to Twitter's policies. We would find these ads during our regular scrapes for ads by all federal candidates, but no billing or impression data would be available, and the ads would disappear from Twitter's archive after 7 days, as is typical for non-political ads. We notified Twitter about 4 accounts which they subsequently added to their transparency archive. However, Twitter did not retroactively include their prior promoted tweets from these accounts and there are currently 11 additional federal candidate accounts which have promoted tweets not included in the archive. Thus, it appears that Twitter's process for flagging federal election candidates' account that should be included in their archive is not working correctly. We will disclose this new set of 11 accounts to Twitter and continue to work with them to improve their process for discovering and including relevant promoted tweets in their transparency archive.

\section{Datasets}

We have collected all of the U.S. political ad data that Google and Twitter have made transparent and archived as of \studyEndDate. In addition, we have made our best effort to collect as much of the U.S. political ad data that Facebook has made transparent and archived as of \studyEndDate. For Facebook, we are not able to collect all of the ad data from their transparency archive due to the limitations in their API; this is a subset of U.S. political ads that ran on Facebook. Note that our scraper was blocked by Facebook in mid-July, 2018 and we were not able to collect data until the beginning of September, 2018 when we began to use their beta API. This means that we do not have good coverage of Facebook ads during that period since it is difficult to retrieve older ads from Facebook's current beta API. On October 23rd, 2018, shortly after we froze our dataset, Facebook released their Ad Archive Report~\cite{FacebookAdArchiveReport}. From this, we know that as of the cutoff date for data analyzed in this study, Facebook had a total of \FacebookReportAds\ ads in their archive, from a total of \FacebookReportSpend\ spent across \FacebookReportPages\ pages. We have captured over 75\% of all ads in the archive, but only 49\% of the pages. 

On the Facebook platform, ads that run without a 'Paid for by' label but are later deemed to be political are removed from circulation and added to the archive. We have been able to find \fbUnvettedSponsorAds\ such ads in the archive, with a total spend of at least \$\fbUnvettedSponsorSpend\ and \fbUnvettedSponsorImpressions\ impressions. It does not appear that Google or Twitter have any mechanism for retroactively marking an ad as political if it is discovered after the fact, and we would encourage them to develop this capacity.

Table~\ref{tab:overall_data} shows all of the data that we have collected from each of the platforms. Most of the political ads in these archives are from late May, 2018 to \studyEndDate\, but there are several older ads from Twitter and Facebook that have been included in their transparency archives. Facebook has the most advertisers, ads, impressions, and spend. However, Facebook also includes many political issue ads in their transparency archive that are not included in Google and Twitter's transparency archives so this is not a fair comparison of political advertising activity across all three platforms. An important difference between the datasets is that, while Facebook and Twitter are publishing breakdowns of impressions on geographic and demographic lines, Google is instead publishing geographic and demographic targetings. These should not be considered equivalent. Below in the analysis section we will present a more accurate comparison of political advertising activity across all three platforms.

Additionally, we use a dataset published by ProPublica of political ads that have been viewed by their users who have installed browser extensions that automatically collected advertisements on their Facebook pages and sent them to ProPublica's servers~\cite{Propublica_Dataset}. Table~\ref{tab:propublica_dataset} provides an overview of this dataset. We were able to connect ads in the ProPublica dataset to ads in our dataset of archived political ads by mapping the ad IDs used in the ProPublica dataset to the ad archive IDs used in the archive. To do this, we scraped the Facebook's web-based political ad archive, as both the ad IDs and ad archive ids were available. Each record in this dataset contains, among other things, the text of the data, the various targetings received by the different users who saw the ad, the page associated with the ad, and the 'Paid by' ad sponsor string associated with the ad. Of the \postArchiveProPublicaAds\ ads in the ProPublica dataset with a creation date after May 7$^{th}$, 2018, the official start of the Facebook dataset, we were able to find \propublicaAdsInArchive. Because the users who contribute data to this dataset are self-selecting, these ads should not be considered a representative sample of ads in the larger Facebook Ad Archive. Among other things, the average ad spend on ads in this dataset was \$\propublicaAverageSpend, compared to \$\averageFBSpend\ for the larger dataset.

As part of our analysis, we manually categorized the top advertisers on all three platforms. We categorized these advertisers by organization type (political candidate, Political Action Committee (PAC), Union, For Profit, etc). For Facebook, we were able to classify the organizations of the advertisers who were responsible for at least 75\% of the total number of ads in the Facebook archive. For Google, we labeled the organization of top advertisers who were responsible for 80\% of the total number of ads, and for Twitter we we were able to label all \overallTwitterAdSponsors\ advertisers with their organization type. We were able to categorize \overallCategorizedAdSponsorCount\ of the top ad sponsors. If we were not able to categorize an advertiser, it is marked as 'Unknown'.

We also classified the ads themselves into 5 categories: Inform, Connect, Donate, Move, or Commercial. Inform ads seek to persuade the viewer, but do not make an explicit ask. Connect ads seek the user's contact information. Donate ads seek the user's money. Move ads attempt to motivate the user to take some action in the physical world, such as attending a rally or voting. Commercial ads seek to sell the user goods or services. We classified the ads based on the outgoing links from the ads. Ads that had no outgoing links were always classified as Inform ads, as they could not have any further ask from the user. Ads that linked directly third-party sites for event management (eventbrite.com), contact management(Google Docs), or payments management (actblue.com) were solely classified as Move, Connect, or Donate ads respectively. Ads that linked to general campaign sites were usually multiple-classed as some combination of the three, as these ads and pages typically made multiple asks. Ads by For Profit Media organizations were classified as Inform ads, as these advertisers do not sell goods or services to users. Ads by For Profit organizations that linked to store sites or sites selling services were classified as Commercial. We were able to categorize \overallCategorizedAdCount\ ads with these methods. Heavy use of third-party service providers by advertisers was extremely helpful in making these classifications. If we were not able to categorize an ad, it was marked as 'Unknown'. We validated this method of ad categorization by taking a random sample of 300 categorized ads from each platform and manually verifying them. The error rate for Facebook was 4\%, for Google was 3.7\%, and for Twitter was 3.7\%

A limitation that applies to all our datasets is that we do not know when the spend and impressions for each ad occurred during the lifetime of the ad. Some ads run for several weeks and some for only a day, but in either case, we attribute their entire spend and total impressions to the creation date of the ad.

\begin{table*}
    \centering
    \begin{tabular}{|l|l|l|l|l|l|l|l|}
     \hline Platform & Total Ads & Total Sponsors & Total Pages & Impressions  & Spend & First Ad Date & Last Ad Date\\ \hline
     Facebook & \overallFBAds & \overallFBAdSponsors & \overallFBPages & \overallFBMinImpressions\ - \overallFBMaxImpressions & \$\overallFBMinSpend\ - \$\overallFBMaxSpend & \overallFBFirstDate & \overallFBLastDate \\ \hline
     Google &  \overallGoogleAds & \overallGoogleAdSponsors & NA & \overallGoogleMinImpressions\ - \overallGoogleMaxImpressions & \$\overallGoogleSpend  & \overallGoogleFirstDate & \overallGoogleLastDate \\ \hline
     Twitter & \overallTwitterAds & \overallTwitterAdSponsors & NA & \overallTwitterImpressions & \$\overallTwitterSpend  & \overallTwitterFirstDate & \overallTwitterLastDate \\ \hline
    \end{tabular}
    \caption{Overall Datasets}
    \label{tab:overall_data}
\end{table*}{}

\begin{table}
    \centering
    \begin{tabular}{|l|l|}
        \hline Total Ads &  \overallProPublicaAds \\
        \hline Total Pages & \overallProPublicaPages \\
        \hline Total Ad Sponsors & \overallProPublicaAdSponsors \\
        \hline Earliest Ad Date & \overallProPublicaFirstDate \\
        \hline Latest Ad Date & \overallProPublicaLastDate \\
        \hline 
    \end{tabular}
    \caption{ProPublica Political Advertisements From Facebook}
    \label{tab:propublica_dataset}
\end{table}

\section{Results}
We calculate total spend and impression minimum and maximum for Facebook ads by summing respectively the smallest and largest value for the range given for each ad. For Google, advertiser weekly spend data was aggregated for all advertisers, so we did not have to estimate that number. For Twitter, exact numbers for impressions and spend were available, so no estimation was needed. We also note that we are only able to collect a subset of political advertisements from Facebook's transparency archive due to accessibility issues with their beta API.
We stress that because the criteria for inclusion in these archives differed on the different platforms, the figures on relative proportions of ad types and advertiser types should be seen as a reflection of what the platforms chose to make transparent in addition to what is organically present on these platforms. 

\begin{figure}
    \centering
    \includegraphics[width=3.25in]{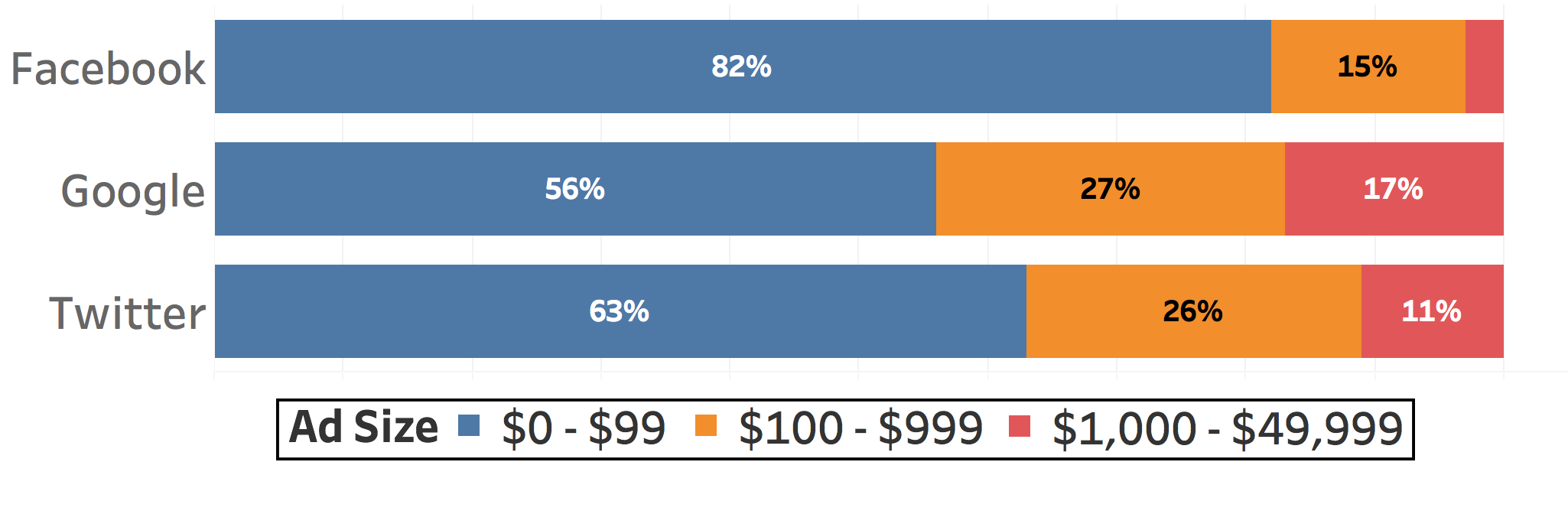}
    \caption{Distribution of ads by size}
    \label{fig:ads_by_size}

\end{figure}

\begin{figure}
    \centering
    \includegraphics[width=3.25in]{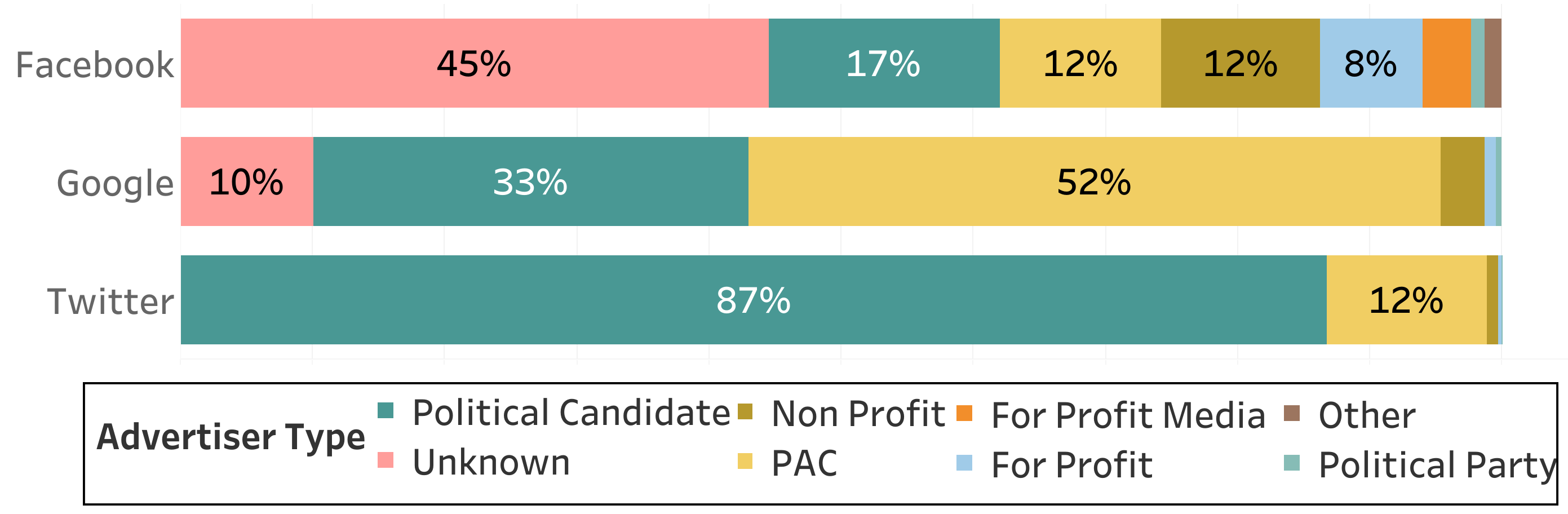}
    \caption{Distribution of spend by advertiser type }
    \label{fig:spend_by_advertiser_type}
\end{figure}

With that in mind, we can see clear differences between the platforms. Of particular note is the difference in ad size visible in Figure~\ref{fig:ads_by_size}, with Facebook having a much larger sized share of the smallest size of ad. Also of note is the differing prevalence of types of advertisers in Figure~\ref{fig:spend_by_advertiser_type}, with PACs making up a much larger percentage of spend on Google compared with the other platforms.

\begin{figure}
    \centering
    \includegraphics[width=3.25in]{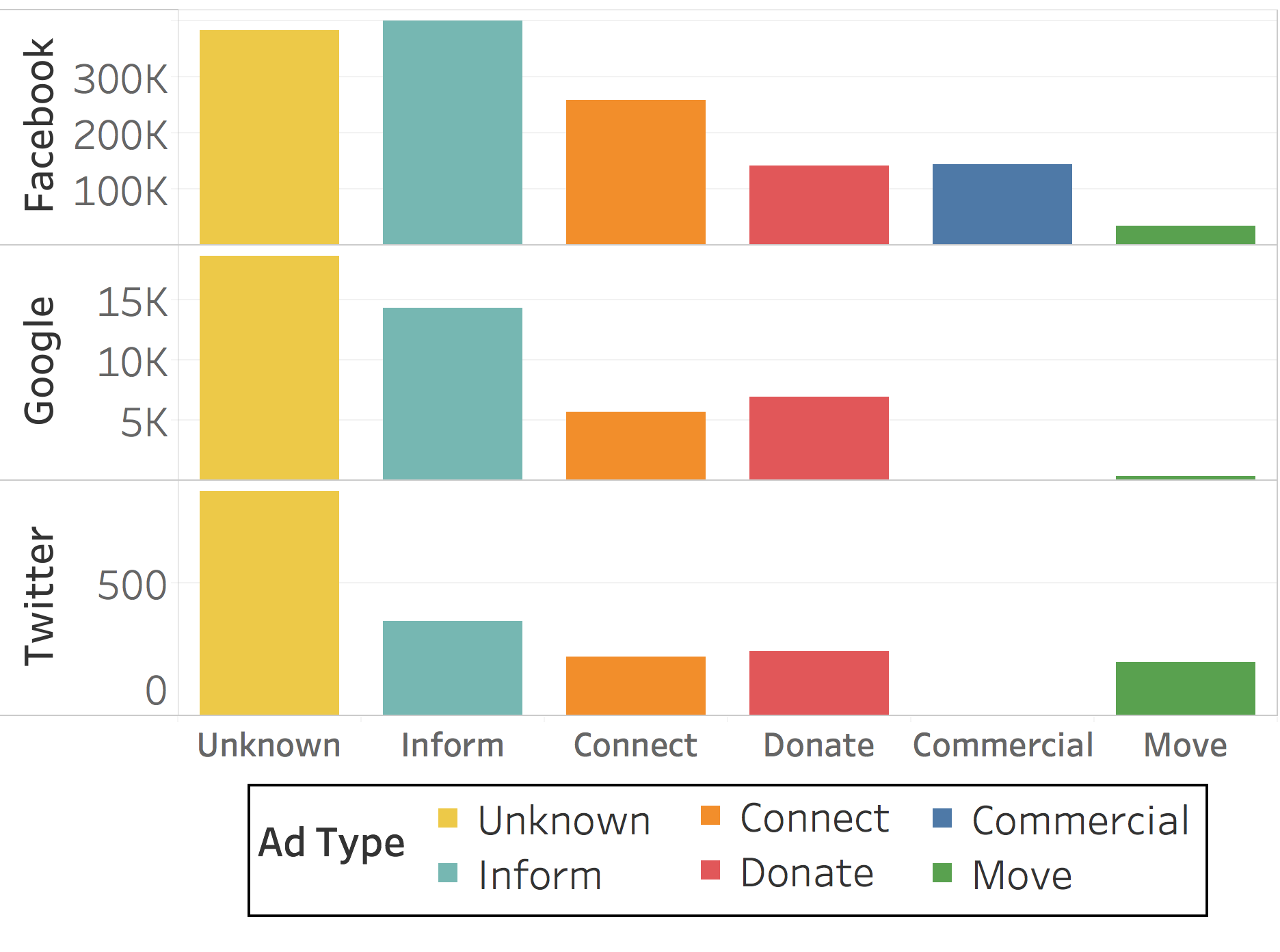}
    \caption{Distribution of ads by type }
    \label{fig:ads_by_ad_type}
\end{figure}

\subsection{Data Over Time}
\begin{figure*}
    \centering
    \includegraphics[width=7in]{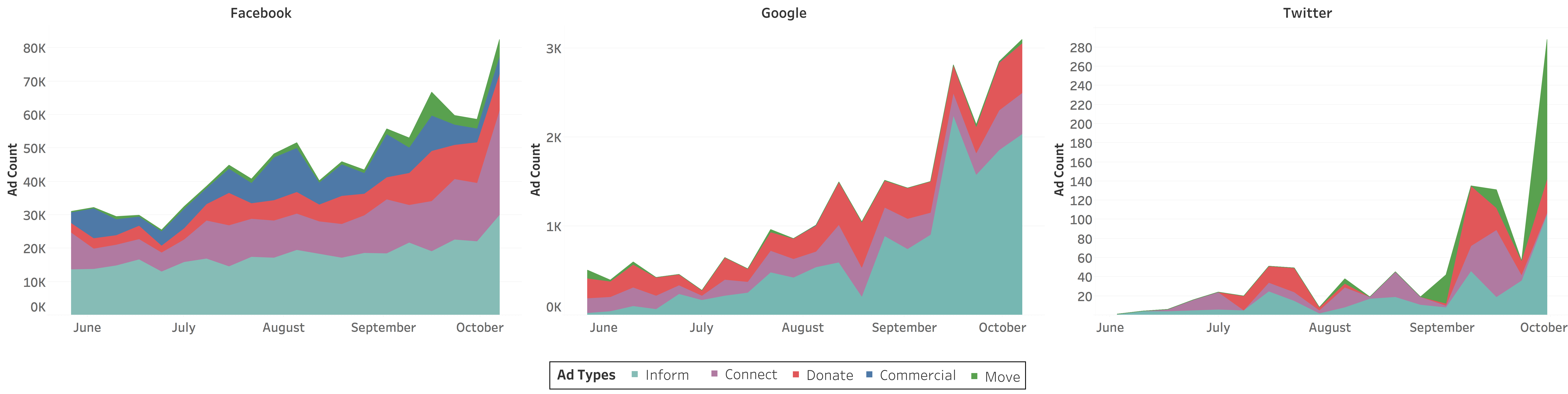}
    \caption{Platform Ad Count By Ad Type By Week }
    \label{fig:_spend_by_week}
\end{figure*}

\begin{figure*}
    \centering
    \includegraphics[width=7in]{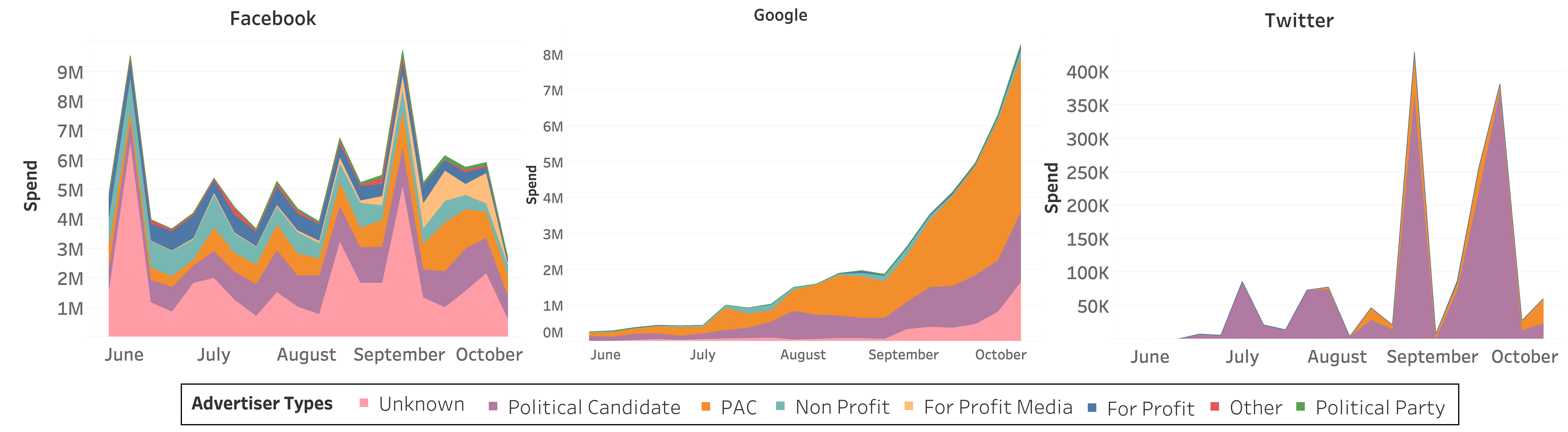}
    \caption{Platform Spend By Advertiser Type By Week }
    \label{fig:_ads_by_week}
\end{figure*}

The time period during which we were collecting data coincided with the 2018 midterm elections in the United States. Thus, we were able to observe changing patterns in spend leading up to a major election. Figure~\ref{fig:_spend_by_week} shows spend by week for the 5 month period leading up to the election and Figure~\ref{fig:_ads_by_week} shows raw ad count for the same period. We note that our data, particularly for Facebook spend, is right-censored for the final two weeks. This is caused by Facebook's API limitations which only enable us to be able to recheck ad spends weekly. Thus, newly create ads have likely not spent much of their budget when we initially discover them. This right censor effect also likely effects Google and Twitter to a lesser degree due to ads with larger budgets that take several days to spend down completely. This can be corrected by periodically rechecking the ads until they have all spent their budgets which is normally within a week. If the paper is accepted, we will update the data to include ads up to the U.S. midterm elections.

We can see the expected increases in the number of ads on all three platforms as the U.S. midterm elections approach. On Facebook's platform there is an increase in connect ads and on Twitter there is an increase in move ads. Both of these are related to sophisticated ``get out the vote'' efforts that many groups have deployed. These move ads include images which include specific polling place addresses and websites that provide polling place directions and information. The connect ads often provide users with instructions on how they can volunteer to help with early and day-of voter turnout efforts. The cause of the spending spikes for Facebook's platform can be attributed to a few unknown sponsors that we could not link to a legally registered entity but that were likely quasi for-profit advertisers which we will discuss further later in the paper. The spending spikes on Twitter's platform can be attributed to candidates who ran a few ads with larger budgets.

\subsection{Federal Candidate Comparison}
In order to understand how political advertising across these platforms differ, we attempted to create a comparable subset of advertisers and ads. This is difficult because each platform has slightly different criteria for inclusion. To do this, we present results for advertising only paid for by candidates for federal office, which was the broadest set that was reliably included in all three archives.  Note, this does not include ads by current officeholders who are not seeking re-election or ads that merely mention a federal candidate but are paid for by another party. Results for these advertisers are presented in Table~\ref{tab:fc_results}. 

\begin{figure}
    \centering
    \includegraphics[width=3.25in]{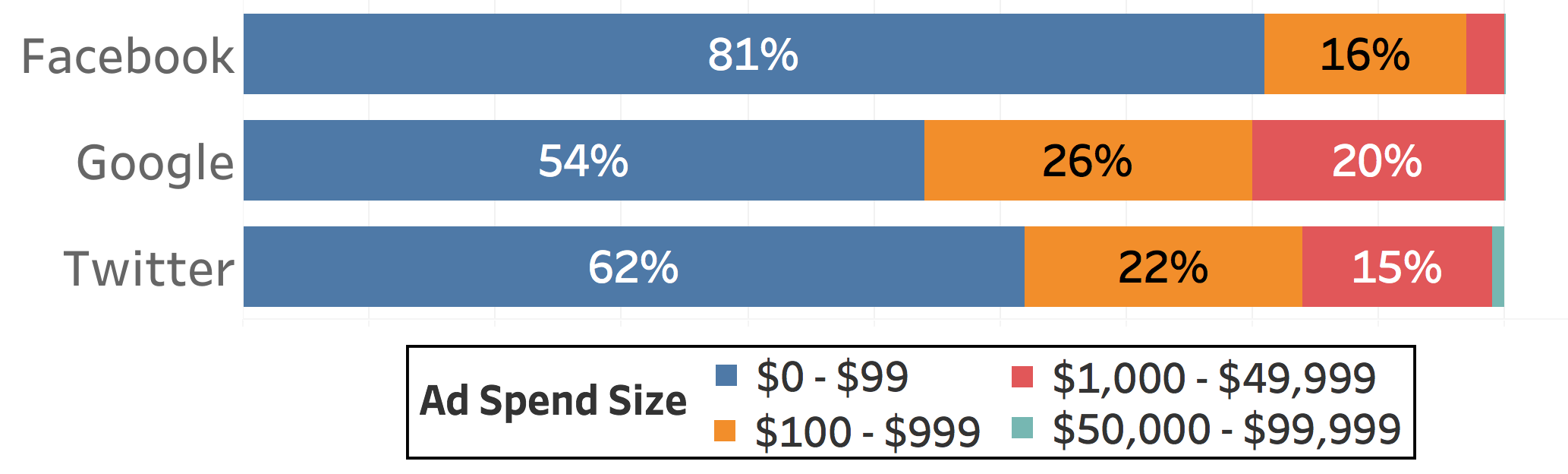}
    \caption{Federal Candidate ads by Size }
    \label{fig:fc_ads_by_size}
\end{figure}

\begin{figure}
    \centering
    \includegraphics[width=3.25in]{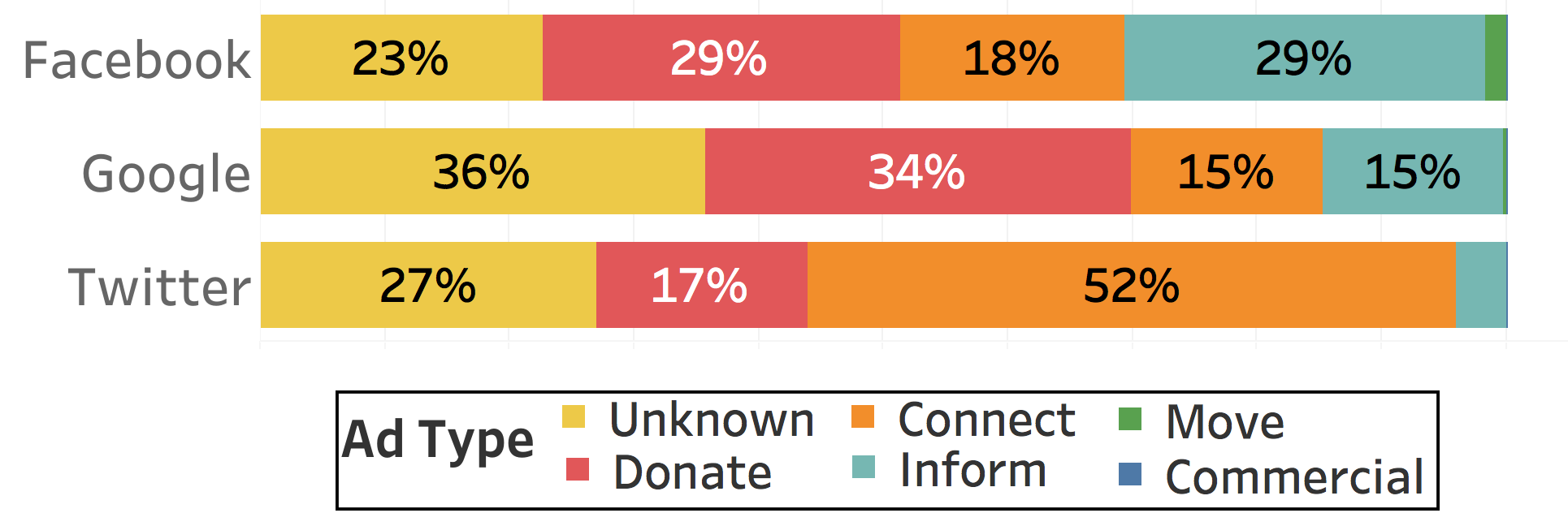}
    \caption{Federal Candidate Spend by Ad Type }
    \label{fig:fc_ads_by_type}
\end{figure}

\begin{figure*}
    \centering
    \includegraphics[width=7in]{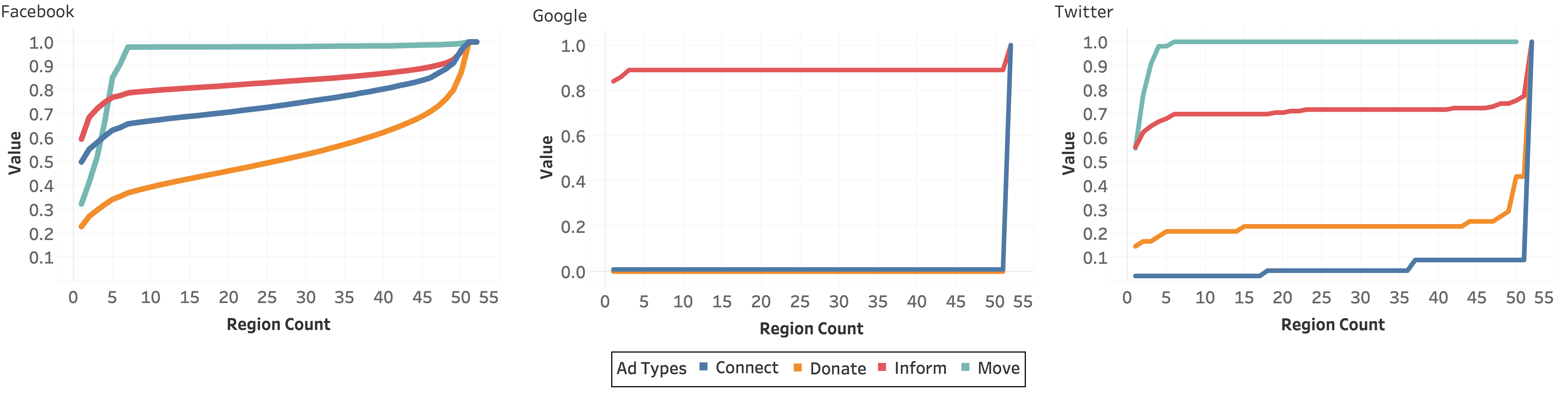}
    \caption{CDF of Regions by Ad Type for Federal Candidates}
    \label{fig:fc_cdf_regions_ad_type}
\end{figure*}

\begin{table}
    \centering
    \begin{tabular}{|l|l|l|l|}
        \hline Results & Facebook & Google & Twitter \\ \hline
        Total Advertisers & \totalFBAdvertisersFC & \totalGoogleAdvertisersFC & \totalTwitterAdvertisersFC \\ \hline

        Total Ads & \totalFBAdsFC & \totalGoogleAdsFC & \totalTwitterAdsFC \\ \hline
        Total Impressions & \totalFBImpressionsFC\ - \totalFBMaxImpressionsFC & \totalGoogleImpressionsFC - \totalGoogleMaxImpressionsFC & \totalTwitterImpressionsFC \\ \hline
        Total USD Spend & \$\totalFBSpendFC\ - \$\totalFBMaxSpendFC & \$\totalGoogleSpendFC & \$\totalTwitterSpendFC \\ \hline
        Ave Impressions/Ad & \averageFBImpressionsFC\ - \averageFBMaxImpressionsFC & \averageGoogleImpressions\ - \averageGoogleMaxImpressions & \averageTwitterImpressions \\ \hline
        Ave USD Spend/Ad & \$\averageFBSpendFC\ - \$\averageFBMaxSpendFC & \$\averageGoogleSpend & \$\averageTwitterSpend \\ \hline
    \end{tabular}
    \caption{Federal Candidate Only Results}
    \label{tab:fc_results}
    \vspace{-1cm}
\end{table}

Table~\ref{tab:fc_results} shows that Facebook is the platform with the broadest appeal to federal candidate advertisers, with far more advertisers and ads than Google. However, political advertising by this group on Google appears to generate more spend and possibly more impressions than ads on Facebook. The average ad size on Facebook in terms of impressions and spend are the smallest based on our minimum estimates indicating that advertisers are running smaller, likely more targeted ads on Facebook. These small ads on Facebook are what are called micro targeted, which we define as less than 1,000 impressions or a spend of less than \$100. For Facebook, microtargeted ads make up 81\% of the overall number of ads for federal candidates in our dataset. For Twitter, this number is 62\%, and for Google it is 54\%. Figure~\ref{fig:fc_ads_by_size} shows the share of ads by size of spend, and here we can begin to see how federal candidates use these platforms in different ways. We note that the distribution of ads by size for federal candidates in Figure~\ref{fig:fc_ads_by_size} is very similar to the overall distribution of ads by size seen in Figure~\ref{fig:ads_by_size}. 

Figure~\ref{fig:fc_ads_by_type} shows the relative spend on different ad platforms, where we see very different percentages for types of ads. Commercial ads are not shown in this figure because there were too few commercial ads to be visible. Particularly of note is the fact that ads seeking donations were far more common on the Google platform, and ads seeking to spread a message ('Inform') were much more common on Facebook. 

Seeing these differences in both ad size and the types of ads that were run, we wanted to understand if advertisers were trying to reach different geographic audiences with different types of ads. To do this, we compared the number of regions in which various ads had impressions on Facebook and Twitter, and the number of regions targeted for Google. Figure~\ref{fig:fc_cdf_regions_ad_type} shows that a variation in targeting strategy is visible on Facebook and Twitter. On Facebook, 'Move' ads that encouraged people to attend a rally, volunteer for a candidate, or some other in-person activity were viewed, on average, in 4 regions, while 'Donate' ads were viewed in 24 regions on average. This makes a certain amount of intuitive sense; people are willing to travel only so far to attend a rally, but can donate to candidates anywhere in the United States.

\subsection{Ad Targeting}
One of the deficiencies with the Facebook political ad archive is that while it does share geographic and demographic information about who saw a particular ad, we have no way of knowing how that ad was targeted. However we were able to connect our dataset containing information about who consumed ads with one published by ProPublica, which contains some data about how ads were targeted. The ProPublica data was collected by a browser plugin operated by ProPublica which anyone can install. ProPublica's browser plugin~\cite{Propublica_Plugin} uses a supervised Natural Language Processing (NLP) classifier to detect political ads in addition to allowing users to manually classify ads they see as political. The browser plugin then collects the partial ad targeting explanations Facebook provides by automatically clicking on the ``Why am I seeing this?'' button for political ads and sends it to ProPublica for them to make public. 

We first provide a brief background on Facebook targeting audience options~\cite{venkatadri-2018-targeting}. Facebook exposes prospective advertisers to a plethora of options. First, advertisers can target users based on age, gender, location and languages they speak. Second, advertisers can choose to send their ads to users in a custom audience or lookalike audience. Custom audiences contain a list of identifiers of specific users. Advertisers can use various types of data to create a custom audience list, ranging from specifying the emails, phone numbers or physical addresses of people they want to reach, to users that have visited their website, installed their mobile application, or liked their Facebook Page. Lookalike audiences allow advertisers to let Facebook choose to whom to sends their ads based on previous campaigns. Finally, advertisers can choose from a long list of targeting attributes the characteristics they want users who receive their ads to have (e.g., users interested in Catholic Church). Targeting attributes are categorized in types such as demographics, behaviors and interests. Advertisers can choose multiple attributes, to target.

A prior study by Athanasioshas, et al.~\cite{EURECOM5414} reverse engineered what Facebook chooses to show and the limitations of the ad targeting explanation Facebook provides. This study showed that ad explanations are incomplete; each explanation, shows at most one targeting attribute (plus age/gender/location information), regardless of how many attributes the advertisers use. This means that explanations reveal only part of the targeting attributes that were used, providing us -- and the users -- with an incomplete picture of the attributes that advertisers were using. However, in the same study, authors performed a number
of controlled experiments that suggest -- but not conclusively prove -- that there is a logic behind which attributes appear in an explanation and which do not. Given a targeting audience A obtained from two attributes $a1$ and $a2$, if $a1$ and $a2$ come from different attribute categories (e.g. Demographic, Behavior, Interest, etc.), the attribute shown follows a specific precedence (Demographics and Age/Gender/Location > Interests > PII based lists > Behaviors). If $a1$ and $a2$ come from the same attribute category, the one that appears in the explanation is the one with the highest estimated audience size. This will result in a systematic under-counting of lower priority targeting types. 

There are two main sources of biases and limitations in ProPublica's dataset. One comes from users that installed ProPublica's plugin and which political ads they were shown. Another is from the way Facebook provides ad explanations. The ProPublica dataset is the only publicly available source of targeting information for Facebook political ads. Thus, we present these results to provide an initial insight into how Facebook political advertisers are targeting their ads with the understanding of likely biases and limitations.

\begin{figure}
    \centering
    \includegraphics[width=3.25in]{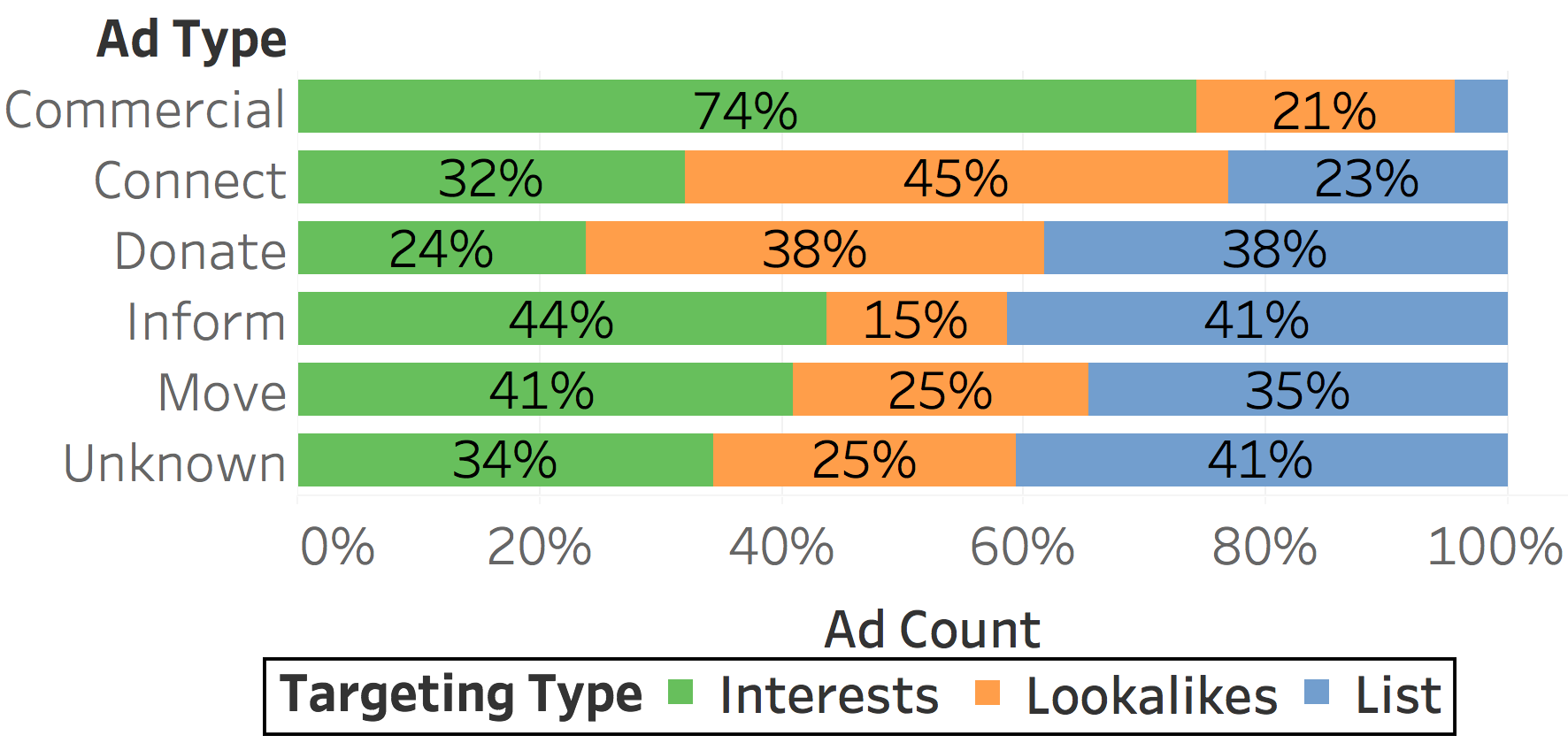}
    \caption{ProPublica Spend by Ad Type }
    \label{fig:pp_ads_by_type}
\end{figure}

With these caveats in mind, we proceed to an analysis of the \propublicaAdsInArchive\ ads which we were able to connect between the ProPublica dataset and ours. In Figure~\ref{fig:pp_ads_by_type}, we see that different types of ads do indeed rely on different targeting strategies. Of particular note is the the divergence of 'Commercial' ads, of which 74\% rely on targeting by interest groups, and of 'Donate' ads, of which only 24\% do. The average ad size did not differ significantly between targeting types, but was significantly larger than the average for the Facebook archive as a whole. We believe this to be an artifact of the collection mechanism, which is biased toward finding larger ads.
91\% of ads in the overall ProPublica dataset had some kind of geographic targeting, and 92\% had age or gender targeting. On average, ads in this dataset had on average 4.1 different targeting parameters, so for Facebook these should be thought of as a minimum criteria. By contrast, 58\% of ads in the Google archive had no geographic targeting whatsoever and 70\% had neither age nor gender targeting.

\begin{figure}
    \centering
    \includegraphics[width=3.25in]{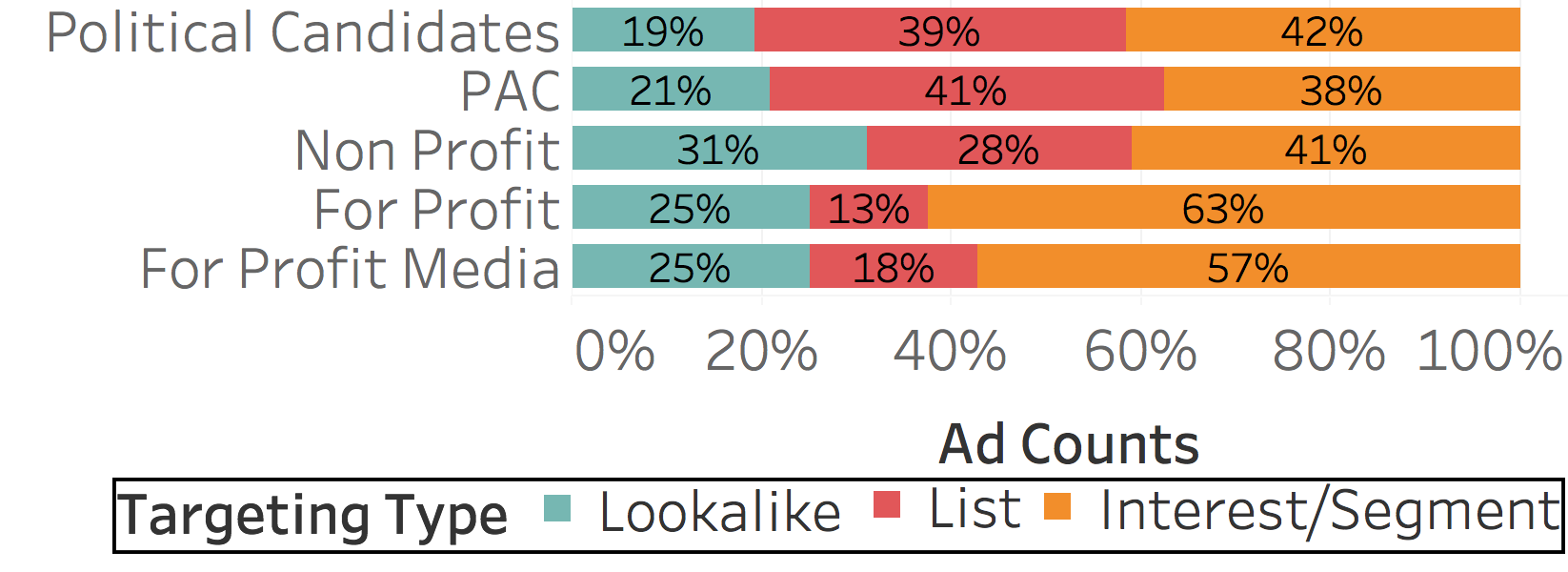}
    \caption{ProPublica Targeting by Advertiser Type }
    \label{fig:targeting_by_advertiser}
\end{figure}

In Figure~\ref{fig:targeting_by_advertiser}, we also see diverging strategies between advertisers, with Political Candidates and PACs making heavy use of custom lists of users and For Profit and For Profit Media companies relying far more on targeting users by their interests. Campaigns have numerous potential sources from which to compile lists of users. In addition to their own lists of donors and voter rolls, campaigns can rent lists from other candidates~\cite{list_rental}.

Both Google and Twitter offer advertisers similar targeting criteria to what we have described for Facebook, including custom audiences and lookalike audiences. Both even allow targeting of users based on interests, although they infer these interests in different ways. Both have made transparent demographic and geographic targeting information for ads in their archive, but without other targeting information, this is an incomplete picture at best. We encourage Google and Twitter to at minimum follow Facebook's example and make transparent to users information about why they have been targetted for ads that they are seeing.

\subsection{New Types of Political Advertisers}

\subsubsection{For-Profit Media}
One advertiser type in particular proved to be an interesting outlier. The category 'For Profit Media' contains advertisers whose ads are not considered traditional news by Facebook (those ads are in a separate part of the archive that we did not include) but have content intended solely to entertain or sway the opinion of the viewer. Over the Facebook dataset as a whole, the average ad sponsor ran ads on \averagePagesPerAdvertiserFB\ pages. Advertisers in the for-profit media category however, ran ads on \averagePagesForProfitMedia\ pages on average. We have examined many of these for-profit media companies to understand why they are running across many Facebook pages. What we have found in numerous instances is unknown for-profit media companies that appear to be creating disingenuous communities that appear to be ``grassroots movements'' to target different demographics and interests with a combination of paid and organic political messaging. 

A good example of this type of advertiser is "New American Media Group LLC". This ad sponsor ran left leaning ads on 10 different pages. These pages were designed to appeal to different demographics ("Melanin" for people of color, "The Soldier Network" for Veterans, "Raising Tomorrow" for parents, etc.) but often run the same content on multiple pages. While this LLC has an extremely similar name to a now-defunct genuine left-leaning media outlet (New America Media), it appears to have no connection to that prior group, and also appears to have no activity off of Facebook.

While some advertisers in this category were fairly traditional entertainment websites (i.e., Comedy Central), some were ``for-profit'' companies in name only that appeared to exist for no other purpose other than to spread a particular political message and had no way of generating an actual profit. We also discovered ``News for Democracy'' is an LLC that ran left leaning ads on 14 different Facebook pages most of which were designed to be appealing to groups with traditionally conservative view points, such as ``The Holy Tribune.'' Journalists investigated this LLC and linked it to MotiveAI which is a liberal political advertising company.

\subsubsection{Corporate Astroturfing}

Corporations paying for political advertising is not an entirely new phenomenon and has traditionally been funded through industry trade groups and PACs. However, the reporting requirements by the FCC for U.S. political advertising on television often made this political messaging traceable to the real sponsor. These stricter reporting requirements do not apply to online political advertising and the ad-hoc reporting requirements that online platforms have enacted are being abused by corporations and industry trade groups to undo transparency efforts.

We discovered in our analysis 355 ads sponsored by ``Citizens for Tobacco Rights'' which is not a registered company in the U.S. but does disclose on their website and Facebook page that it is operated by cigarette company Philip Morris. However, someone who only saw the Facebook ad disclaimer would not be able to connect the ads to Philip Morris without further investigation. Other journalists have found instances of oil and insurance lobbying groups that also provided sponsor names that did not match the legally incorporated entity sponsoring the ads~\cite{Merrill}. These organizations are seemingly taking advantage of Facebook's policy of not vetting sponsor names since some of these entities also ran political ads on Google's ad platform but provided Google with their EIN (tax ID) and correct legally incorporated names of their organizations~\cite{Merrill}.

\subsection{Discussion}

The different policies, bugs, idiosyncrasies, and security weaknesses of each transparency archive implementation present challenges to our analysis efforts. We find many of the issues with these archives likely stem from a combination of their hasty creation and the fact that the platforms are still working out how to improve security of these archives such they are difficult to deceive or evade. We will first discuss issues related to accidentally or intentionally deceiving these transparency efforts and how they might be improved by implementing more robust sponsor attribution techniques. The second part of our discussion will focus on issues related to bypassing inclusion into the different platforms' archives and what can be done to improve these issues.

\subsubsection{Sponsor Attribution}

The for-profit political advertisers appear to be the ones that are accidental or intentionally skirting and violating the spirit of online transparency sponsorship disclosure policies. As we discussed in the Ad Targeting section, it's extremely easy for groups such as 'New American Media' to  obscure who they actually are from users and researchers.

It is worth noting that such advertising by for-profit corporations was not legal until the Citizens United Supreme Court decision in 2010~\cite{Citizens_United}  that struck down restrictions on election spending by for-profit corporations. However, political messaging advertisers who run ads on television or radio stations governed by the FCC must still report the name and contact information of the business which paid for the ad, including the company's officers and directors. Such data is published by the FCC in a public database. Political advertisers who send direct mail through the U.S. Postal Service (USPS) must also report their activities through the FEC with similar public disclosure of the name and contact information of the business. The regulations that require such disclosure for ads that mention candidates do not apply to online advertising, largely because the laws that mandate such public disclosures were drafted before these platforms were as ubiquitous as they have become. 

What this means in practice though, is that people who want to publicize a political message can form a for-profit company for doing so with no intent of making an profit. As a private company, they do not need to publicly disclose their investors in the way that PACs are required to disclose their donors. Then, the for-profit company can advertise on social media, also without disclosing the legal entity providing the funds to pay for the ad. 

On Facebook's platform, advertisers can easily mislead when providing the 'ad sponsor' string associated with their ads, either intentionally or accidentally. Thus, it is effectively free to circumvent Facebook's transparency implementation. We see numerous instances on Facebook's platforms of this occurring. Sometimes, the unreliability of the ad sponsor label appeared to be caused purely by human error, such as typos or variation during data entry. For example, Donald J. Trump For President, Inc. sponsored ads on both the Donald J. Trump page and the Mike Pence page. However, when sponsoring ads on the Donald J. Trump page, the organization is known as 'Donald J. Trump For President, Inc' and when sponsoring ads on the Mike Pence page, is known as 'Donald J. Trump For President, Inc.'. Facebook has not publicly stated plans to implement additional vetting of political sponsors. Facebook's argument is that anything they might implement for additional vetting would not be scalable because of their broader inclusion policy which extends to political issue ads~\cite{Merrill}. However, this has created a weakness in Facebook's transparency implementation that greatly diminishes its effectiveness for studying dishonest political advertisers.

Google and Twitter both vet sponsors so companies must either reveal their legally incorporated name, pay existing third-parties to create ads on their behalf, or create shell organizations (i.e., LLC, PACs). We should note that we see instances of political ads on Facebook and Twitter where the sponsor is a third-party advertising agency instead of the actual entity that paid for the ads. This is an example of the complexities of correctly attributing political ads to the real sponsors. It is clear from analysis that we need more discussion about how to implement sponsorship disclosure and vetting in a way that makes it practical to deploy at scale and more difficult to circumvent. 

\subsubsection{Transparency Infrastructure}
As we have noted, we appreciate the speed with which these transparency archives were created. However, the lack of full integration of these archives into the broader ad platforms of these companies is currently hurting the efficacy of these transparency efforts. 

We believe that there are ads on the Google and Twitter platforms that would be considered political content that are not included in their transparency archives because their criteria for inclusion are too narrow or their mechanisms for finding this content are insufficient. More research needs to be done into exactly what the general population considers to be political advertising. We would encourage these platforms to create policies and enforcement mechanisms that will make transparent advertising content that the general population would consider political.

We also encountered several technical and policy issues with the archives as they currently exist. Many ads, particularly in the Google archive, were missing content information. Information on spend and impressions were only available in broad ranges from Facebook and Google. No targeting information or very little targeting information was available from any of the platforms. Facebook required us to sign an NDA that prohibited us from sharing our raw data even with other researchers or even discussing our findings directly with non- U.S. Persons.

We call on these organizations to re-architect their platforms and policies to support full transparency of all political ads. We realize that making the changes we recommend will require investment of time and money both in the technology of these platforms and the corporate culture of the organizations that own them.


\section{Related Work}
\subsection{Online Advertising}

Korolova~\cite{5693335} was the first to point out privacy attacks based on micro-targeted online ads. Followup work has reverse-engineered the targeting options provided by major online ad networks~\cite{8418598} and explored privacy~\cite{EURECOM5414} and bias~\cite{pmlr-v81-speicher18a} issues of these online ad networks. There has also been work on designing improved ad transparency mechanisms~\cite{Liu:2013}. For our study, we leverage this prior work on reverse-engineering online advertising networks' targeting options and how Facebook's ad targeting explanation likely is implemented.

To the best of our knowledge, there has been no systematic analysis of online advertisers to this point likely due to the difficulty of collecting large-scale data from online ad networks~\cite{Guha:2010}. One of the only prior large-scale quantitative studies of online advertisers focused on how their strategies effected conversion rates based on aggregate analysis of advertisers on Microsoft's ad network~\cite{Vattikonda:2015}. XRay~\cite{lecuyer:hal-01100757} and Sunshine~\cite{Lecuyer:2015} are two techniques that were created to detect and infer online ad targeting methods. However, these were proof of concept systems and not deployed at large-scale. An initial analysis of Facebook's proposed ad transparency archive implementations pointed out the issue of only including political ads and not revealing targeting information~\cite{upturn}. This report was released before Facebook implemented their transparency archive and therefor did not analyze the ad data archived by Facebook or issues with the actual implementation. We have conducted the first large-scale analysis of online political advertising based on the data recently made transparent by Facebook, Google, and Twitter.

\subsection{Political Advertising}

Analysis of political television ads has been the focus of most prior political advertising studies likely due to this data being publicly published by the FCC and easy to access~\cite{doi:10.1177/0002764205279421,tvAds}. There is at least one prior study that explored the influence of political television ads on online discussion~\cite{doi:10.1111/j.1460-2466.2007.00363.x}. There have also been studies of investigating the polarization of online political discourse~\cite{doi:10.1111/pops.12394,doi:10.1177/0956797615594620}.
The closest related to our study is a prior study which showed that uploading political video advertisements to YouTube generated unpaid organic-views and improved their effectiveness~\cite{youtube}. However, to the best of our knowledge ours is the first large-scale study of online political advertising.

\section{Conclusions}
We have performed an analysis of the ads that we were able to collect from Facebook, Google, and Twitter's transparency archives related to U.S. politics. Based on the data we collected, we provide an initial understanding and taxonomies of online political advertising strategies for both honest and possibly dishonest U.S. political advertisers. We also point out limitations and weaknesses of the policies and current implementations of these archives. As part of our analysis, we demonstrate how advertisers are intentionally or accidentally deceiving and bypassing these political transparency archives. We provide a concrete list of suggestions that would likely make them more robust and useful for enabling a better understanding of political advertising. We are actively working with each archive product teams to improve their implementations.

We commend Facebook, Google and Twitter for their efforts so far in improving transparency into political advertising on their platforms. We note the speed with which these archives were made available after public concern about this issue was raised, and that these transparency efforts have improved a great deal in the short time that these tools have been available. We encourage the platforms to continue to improve.

